# UBIQUITOUS MOBILE HEALTH MONITORING SYSTEM FOR ELDERLY (UMHMSE)

Abderrahim BOUROUIS[1], Mohamed FEHAM[2] and Abdelhamid BOUCHACHIA[3]

[1]STIC laboratory, Abou-bekr BELKAID University, Tlemcen, Algeria
`a_bourouis@mail.univ-tlemcen.dz`
[2]STIC laboratory, Abou-bekr BELKAID University, Tlemcen, Algeria
`m_feham@mail.univ-tlemcen.dz`
[3]Research Group, Software Engineering and Soft Computing, University of Klagenfurt, Austria
`hamid@isys.uni-klu.ac.at`

## ABSTRACT

*Recent research in ubiquitous computing uses technologies of Body Area Networks (BANs) to monitor the person's kinematics and physiological parameters. In this paper we propose a real time mobile health system for monitoring elderly patients from indoor or outdoor environments. The system uses a bio-signal sensor worn by the patient and a Smartphone as a central node.*

*The sensor data is collected and transmitted to the intelligent server through GPRS/UMTS to be analyzed. The prototype (UMHMSE) monitors the elderly mobility, location and vital signs such as Sp02 and Heart Rate. Remote users (family and medical personnel) might have a real time access to the collected information through a web application.*

## KEYWORDS

*Ubiquitous health monitoring, Mobile Health Monitoring, Smartphone. Intelligent central sever, Location.*

## 1. INTRODUCTION

Rising health care costs and an increasing elderly population are placing a strain on current health care services. Elderly patients, particularly those with chronic conditions, require continuous long-term monitoring to detect changes in their condition as early as possible [1].
According to the Algerian National Office Of Statistics [2], the percentage of the total population of persons over the age of 65 has increased and is expected to increase further, the number reached 2.7 million out of a population of 35.5 million Algerians and it will reach 20 percent in 2030. In general, the greater part of elderly suffer from various chronic diseases, based on World Health Statistics (WHO) and other sources, chronicle diseases and psychological pressures are behind the death of 80 percent of elderly people in Algeria [3].
In recent years, initiatives have been taken both from academia and by the industries with a view for improving the health care and safety of the public by taking use of information and communication technologies. Most research activities have been focused on achieving common platform for medical records, monitoring health status of the patients in a real-time manner, improving the concept of online diagnosis, enhancing security and integrity of the patients, developing or enhancing telemedicine solutions, which deals with remote delivery of health care services applying telecommunications, etc [4, 5].
Advances in sensor technology have enabled the development of small, lightweight medical sensors that can be worn by the patient while wirelessly transmitting data. This frees the patient from the confinement of traditional wired sensors, allowing him or her to move at leisure and increasing comfort in daily environment. It is foreseen that with the help of these enhanced mobile health systems, better health care and services can be delivered to users, and hospitals can also benefit a better information management and administration. Also, it will provide the users the ability to access their medical records anywhere and anytime.

Some advantages of using wearable systems to measure mobility are direct access to biomechanical parameters, data logging and processing can be done anywhere, and technological advances are leading to a reduced size, weight, and cost [4, 6].

Compared to laboratory-based systems, wearable technologies take less setup time since multiple sensors and equipment do not have to be attached to the subject and software applications do not need to be started for every session [7]. Recent technological advances in wireless communications, sensor miniaturization and Smartphone processing power, offer great potential in the development of wearable systems for mobility monitoring.

An important area of the mobile healthcare service is the mobile monitoring of the patient's vital signs outside the clinical environment. Mobile healthcare can monitor common vital signs such as blood pressure, electrocardiogram (ECG), pulse rate, blood oxygenation (SpO2), breathing rate, body temperature, body activity and weight, and other measures; this could also be useful in management of chronic disorders and to provide feedback about someone's health in the form of behavioral feedback in order to prevent diseases [8].

In this paper, we describe a new Mobile Health Monitoring system architecture for elderly patients, it uses a wireless body area networks (WBAN) to collect and send data to the intelligent server through GPRS/UMTS.

The rest of the paper is organized as follows: Section 2 provides the short descriptions of the related works; in Section 3 we describe the overall architecture of the UMHMSE system and the functions of major components. In section 4, the implementation of UMHMSE is presented. Section 5 summarizes and concludes the paper.

## 2. RELATED WORKS

The first projects that introduced the mobile health monitoring is presented in [9], MobiHealth project is a health service platform based on a mobile phone as a base station for the wireless sensors worn on the body. It forwards their measurements wirelessly using UMTS or GPRS to a service centre, it provides three services: collecting and storage of the received data, forwarding of data to a doctor or medical centre, and analysis of the data received and the sending of feedbacks to a predefined destination using SMS.

Choi et al. [10] proposed a system for ubiquitous health monitoring in the Bedroom via a Bluetooth Network and Wireless LAN, the system uses Bluetooth and wireless LAN technologies, information gathered from sensors connected to the patient's bed is transmitted to a monitoring station outside of the room where the data is processed and analyzed.

Using the technologies of wireless body area networks (WBAN), Jovanov et al. [7] presented a Wearable health systems using WBAN for patient monitoring. The first level consists of physiological sensors, second level is the personal server, and the third level is the health care servers and related services.

Another example is the WiMoCA from Farella et al. [11] that is a custom-made WBSN where the sensing node consists of a triaxial integrated MEMS (micro-electro-mechanical system) accelerometer. The WiMoCa system's ability has to handle diverse application requirements such as posture detection system, bio-feedback application, and gait analysis.

Morón et al. [12] presented a mobile monitoring system, which can provide medical feedback to the patients through mobile devices based on the biomedical and environmental data collected by deployed sensors, sensors compiles information about patient's location and health status. These data are encrypted to be sent to a server through the mobile communications networks, the system provides access to patient's data, even from a smart phone by a J2ME application.

Dai et al. [13] designed a wireless physiological multi-parameter monitoring system based on mobile communication networks; this system monitors vital signs such as ECG, SP02, body temperature and respiration. Data is transmitted via mobile communications networks to a mobile monitoring station and then to the hospitals central management system where, again, the data must be reviewed and interpreted by a physician or other medical personnel.

Lee et al. [14] presented a ubiquitous monitoring system using the ZigBee protocol to wirelessly transmit patient sensor data so that it may be monitored at a local health station. This sensor data is then transmitted through the internet to a local monitoring station where it is processed and analyzed by a reasoning server.

## 3. SYSTEM ARCHITECTURE

UMHMSE is designed to monitor the elderly: His mobility, location, health status and fall incidents. The system contains three following components (figure 1):

1. Wireless Wearable Body Area Network (WWBAN)
2. Intelligent Central Node (ICN)
3. Intelligent Central Server (ICS)

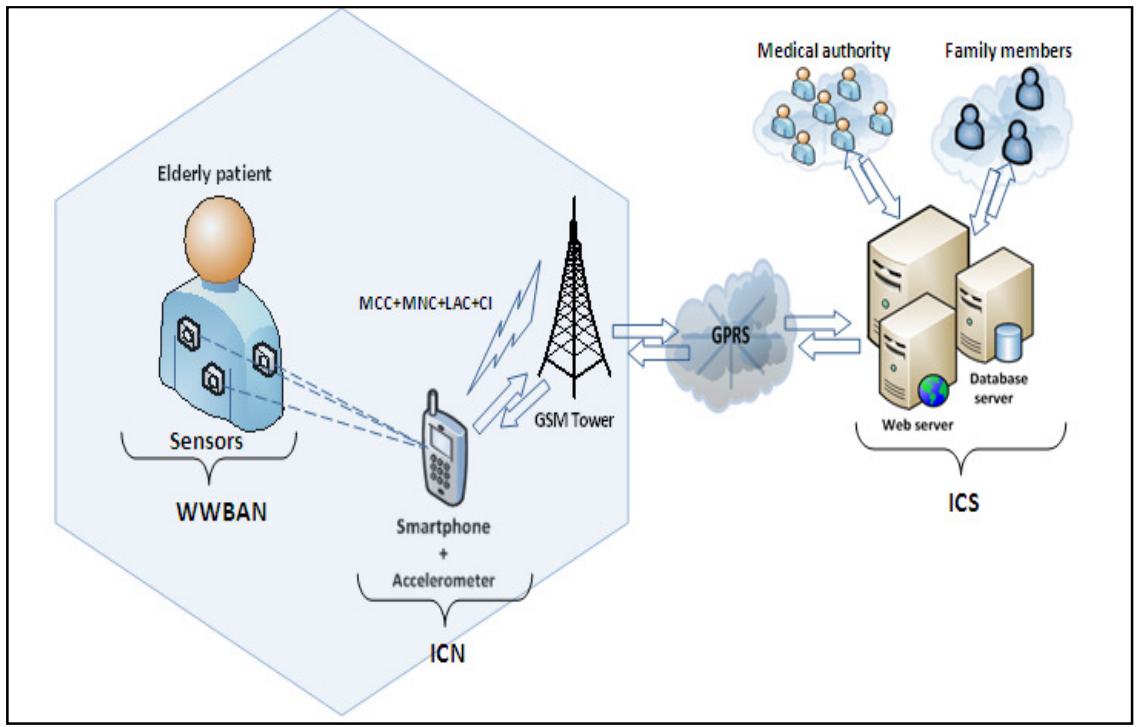

Figure1. UMHMSE System Architecture

### 3.1. Wireless Wearable Body Area Network (WWBAN)

WWBAN consist of one or multiple sensors put and adapted to the body of patient ; the sensors gather their appropriate data and transmit that information to the second component (ICN) via Bluetooth communication protocol.

WWBAN is based on the star topology which implies a centralized architecture where the intelligence of the system is concentrated on a central node which is superior to the peripheral sensors in terms of resources such as processing, memory, and power (figure 2) [15].

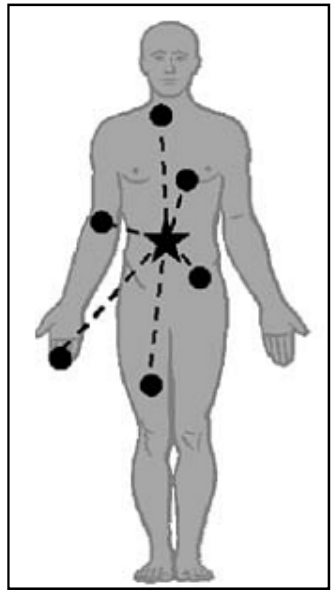

Figure2. Star based WWBAN

## 3.2. Intelligent Central Node (ICN)

ICN is responsible to collect and process the data generated by the WWBAN sensor nodes, UMHMSE uses Smartphone with operating system as the intelligent central node (ICN). It communicates with the ICS using GPRS/UMTS.

With the constant increase in processing power, allowing for sophisticated real-time data processing, Smartphone is a great choice as a central node and body gateway [16]. Other advantage is that Smartphone is often already integrated with sensors; such as accelerometer to determine mobility and global positioning system (GPS) for location, which makes them attractive for a fully integrated wearable mobility monitoring system.

We use also ICN to collect Location Area Identifications such as: Mobile Country Code (MCC), Mobile Network Code (MNC), Location Area Code (LAC) and cell identification (CI) from GSM network to determine the location of the elderly after be calculated in the ICS.[17]

ICN uses a comparison algorithm in order to determine whether to send the information to the ICS or not to save cost for the patient. Sensors Data must be sent when there is a change in collected parameters (health status, mobility and location) figure 3.

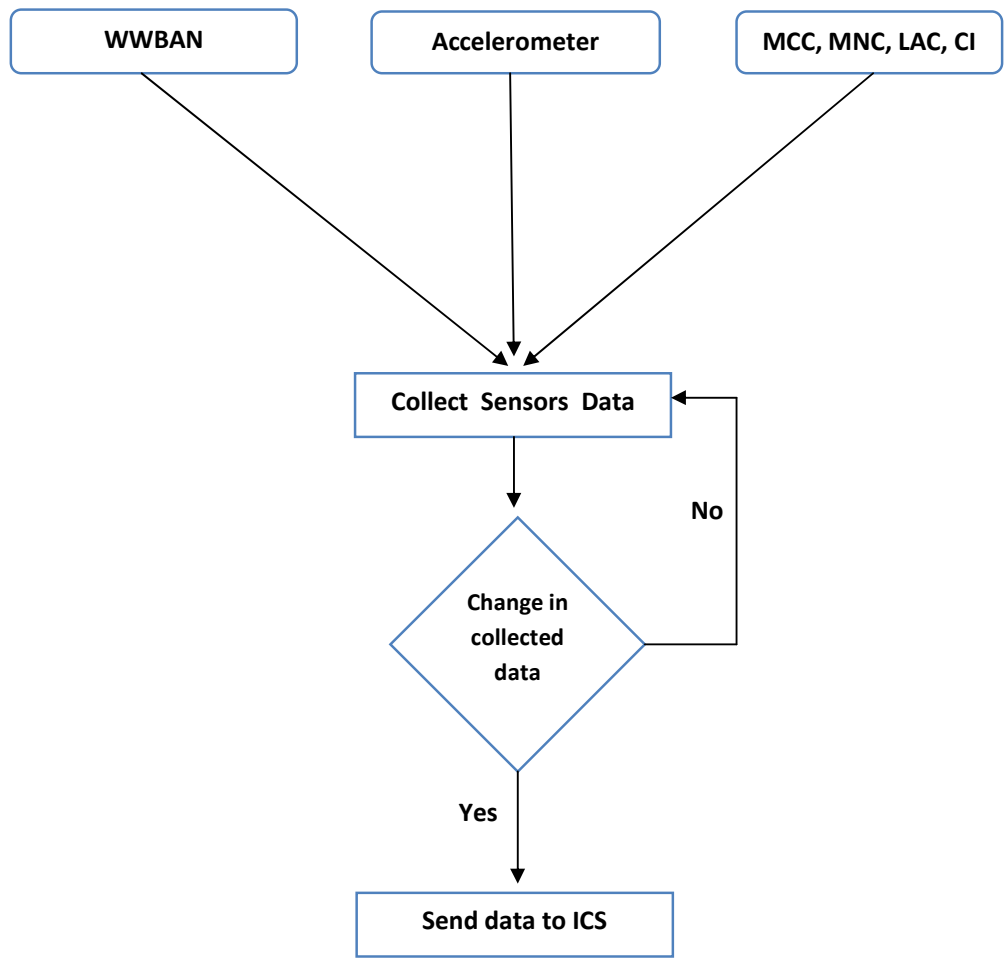

Figure 3. Flow diagram of the ICN function

### 3.3. Intelligent Central Server (ICS)

Intelligent Central Server (ICS) receives sensors data from all the ICN. Once the data is uploaded to the server, it is stored in elderly's database to be analyzed. This analysis is performed autonomously, without human's intervention, comparing the patient's vitals against pre-existing knowledge of his/her condition as well as any recommendations prescribed by the patient's doctor or healthcare professional.

Our system uses logistic regression technique in ICS to mine data and predict health risk from knowledge of the patient's mobility, location and bio-signal sensor data. ICN also uses the combination of input parameters of GSM location (MCC, MNC, LAC, CI) and input body sensors data, because it was noticed that there is a direct factor between health status and place of the Algerian elderly[1].

After processing the information, family members or medical authority can identify the real-time positions and health status of the elderly through a web application. Once an abnormal situation is detected, an alert signal is sent to both give the medical professional an idea of the patient's health status and to alert him of the patient's current location in the case of an emergency.

---

[1] Association of elderly people-Tlemcen-Algeria

## 4. IMPLEMENTATION

We have implemented a prototyping system of UMHMSE. The current sensor being used in the Wireless Wearable Body Area Network (WWBAN) component of UMHMSE is the Nonin 4100 Bluetooth pulse oximeter. It measures blood-oxygen saturation levels (Sp02) as well as heart rate (HR) [18].

We have considered a Nokia Smartphone N95 as hardware platforms of the intelligent central node (ICN) with operating System Symbian v 9.2 based on S60 interface[19]. We developed a Python application for S60 in ICN. Python for S60 is a powerful programming language,it has efficient high-level data structures and a simple but effective approach to object-oriented programming. It an ideal language for scripting and rapid application development in many areas on most platforms [20,21]. Our Python application gathers the sensor data from Nonin sensor (SpO2, HR) and Smartphone integrated sensors (Accelerometer ,MCC ,MNC ,LAC ,CI). ICN uses Python APIs (Application Program Interfaces) to manage BT connections. Once data is received, ICN uses algorithms to compare currents data with the previous; if it detects changes it forwards that data on to the ICS (intelligent central server), figure 4.

ICS includes MySQL database and Apache Server web. All server side functionality is implemented in HTML and PHP (figure 5). Client (ICN) communicates with the server (ICS) using HTTP protocol.

The using of Open Source Software (OSS) (MysSQL,PHP,Apache,Python for S60) implementation will reduce the implementation cost of UMHMSE. [22]

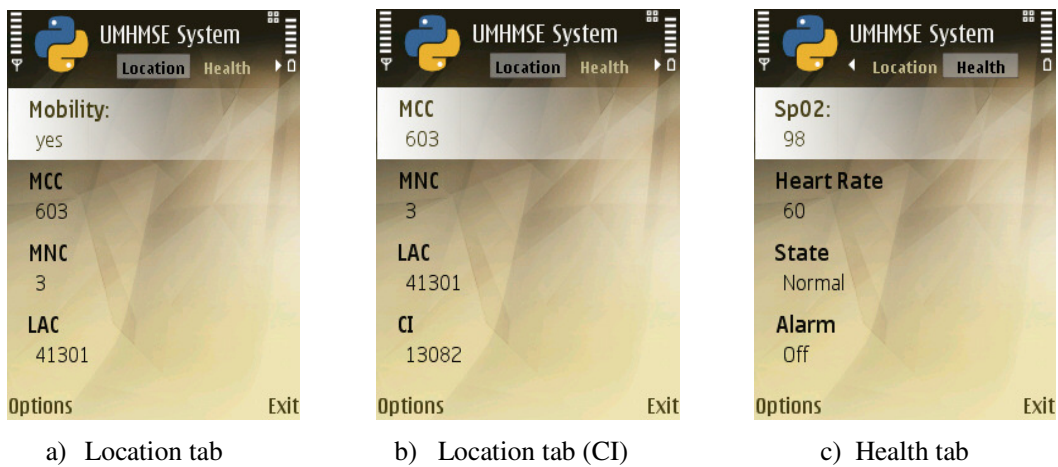

   a) Location tab            b) Location tab (CI)           c) Health tab

Figure 4. Screenshot of the ICN implementation based on Python application

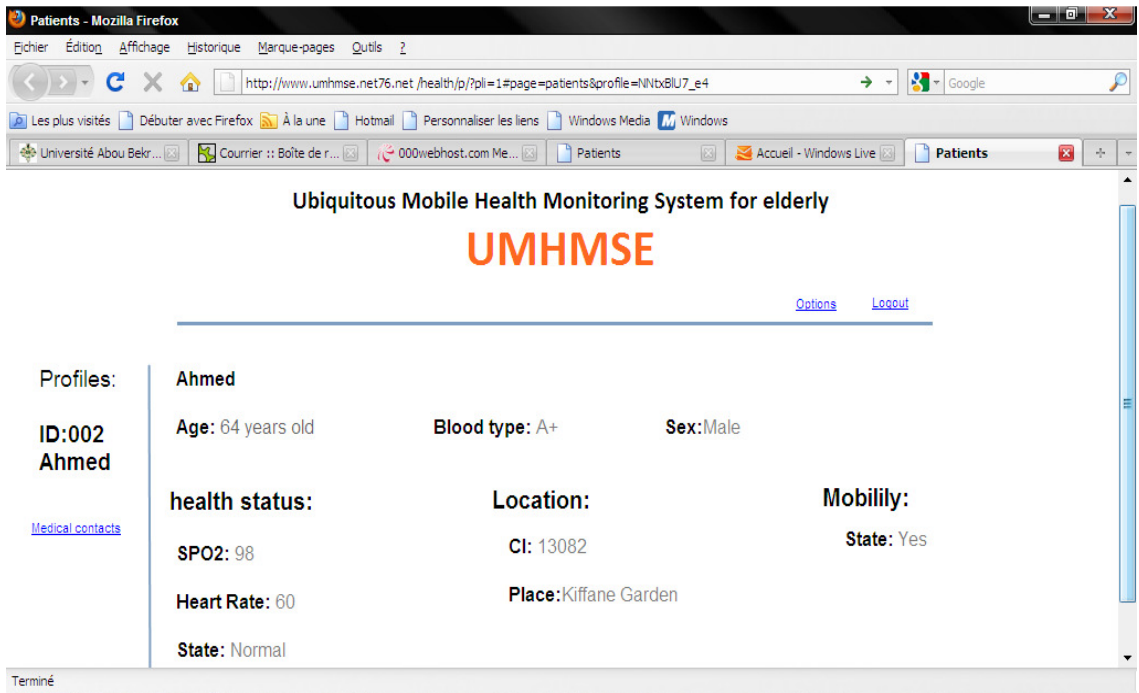

Figure 5. Web interface of UMHMSE

## 5. CONCLUSION

In conclusion, a ubiquitous mobile health system is presented for continuous monitoring of elderly patients from indoor and outdoor environments. The system provides the architecture for collecting, gathering and analyzing data from a number of biosensors. UMHMSE is capable of remotely monitoring human vital signs, mobility and location; it integrates Wearable Wireless Body Area Network (WWBAN), intelligent Central Node (ICN) and Intelligent Central Server (ICS). The designed prototype system monitors location and health status through the use of a Nonin 4100 Bluetooth pulse oximeter as WWBAN and Nokia N95 based on Python application as ICN.By providing capabilities for processing of the measurements and user I/O, the family members or medical authority are alerted in a timely manner when the state of elderly's health changes for the worse.

## ACKNOWLEDGEMENTS

We would like to thank the STIC Laboratory personal Pr. Mohamed FEHAM for their publication support.

International Journal of Computer Science & Information Technology (IJCSIT), Vol 3, No 3, June 2011

## Authors

**Abderrahim Bourouis** received the B.E. and M.E..degrees in telecommunication from Abou-bekr BELKAID university , Algeria, in 2007 and 2009 respectively. He joined STIC laboratory in 2010.  He has been engaged in the design and development of Location-based service (LBS) and Body Sensor Networks (BSN).

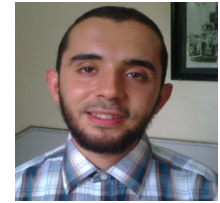

**Mohammed Feham** received  the Dr. Eng. degree in Optical and Microwave Communications from the University of Limoges (France) in 1987, and his PhD in Science from the University of Tlemcen (Algeria) in 1996. Since 1987, he has been Assistant Professor and Professor of Microwave, Communication Engineering and Telecommunication Networks. He has served on the Scientific Council and other committees of the Electronics and Telecommunication Departments of the University of Tlemcen. His research interest now is mobile networks and services.

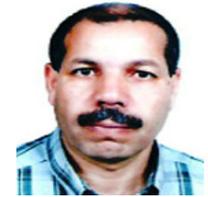

**Abdelhamid Bouchachia** is currently an Associate Professor at the University of Klagenfurt, Department of Informatics, Austria. He obtained his Doctorate in Computer Science from the same University in 2001. He then spent one year as a   post-doc at the University of Alberta, Canada. His major research interests include soft computing and machine learning encompassing nature-inspired computing, neurocomputing, fuzzy systems, incremental learning, semi-supervised learning and uncertainty modeling.. He is a member of the IEEE task force for adaptive and evolving fuzzy systems and member of the Evolving Intelligent Systems (EIS) Technical Committee of the Systems, Man and Cybernetics (SMC) Society of IEEE.

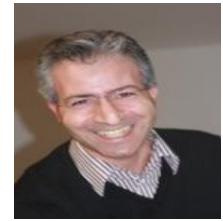